\documentclass[twocolumn,secnumarabic,amssymb, nobibnotes, aps, pre,superscriptaddress]{revtex4-1}

\usepackage{graphicx}
\usepackage{epstopdf}
\usepackage{amsmath}
\usepackage{float}
\begin{document}

\title{Finite-strain Bloch wave propagation by the transfer matrix method}%

\author{M.I. Hussein}
\email[E-mail me at: ]{mih@colorado.edu}
\affiliation{Department of Aerospace Engineering Sciences, University of Colorado Boulder, Boulder, CO 80309-0429, USA}
\author{R. Khajehtourian}
\affiliation{Department of Aerospace Engineering Sciences, University of Colorado Boulder, Boulder, CO 80309-0429, USA}
\author{M.H. Abedinnasab}
\affiliation{Department of Mechanical Engineering, Sharif University of Technology,Tehran, 11365-9567, Iran}
\date{\today}

\begin{abstract}
The introduction of nonlinearity alters the dispersion of elastic waves in solid media. In this paper, we present an analytical formulation for the treatment of finite-strain Bloch waves in one-dimensional phononic crystals. Considering longitudinal waves, the exact dispersion relation in each homogeneous layer is first obtained and subsequently used within the transfer matrix method to derive an approximate dispersion relation for the overall periodic medium. The result is an amplitude-dependent elastic band structure that may be used to elucidate the interplay between the waveform steepening and spreading effects that emerge due to the nonlinearity and periodicity, respectively. For example, for a wave amplitude on the order of one eighth of the unit-cell size in a demonstrative structure, the two effects are practically in balance for wavelengths as small as roughly three times the unit-cell size.  

\end{abstract}

\maketitle

\section{Introduction}
\subsection{Phononic materials}
\textit{Phononic materials} are elastic materials with prescribed phonon wave propagation properties. While the term \textquotedblleft phonon\textquotedblright\,  is formally used in the physical sciences to describe vibration states in condensed matter at the atomic scale, in the present context we use it to broadly describe elastic wave propagation modes. Like crystalline materials, a phononic material has local intrinsic properties and is therefore mathematically treated as a medium that is spatially extended to infinity. Compared to a homogeneous and geometrically uniform elastic continuum, a phononic material exhibits rich and unique dynamical properties due to the presence of some form of non-homogeneity and/or non-uniformity in either an ordered or disordered manner. In the ordered case, phononic materials are constructed from a repeated array of identical \textit{unit cells} which enables the calculation of the elastic band structure for a given topological configuration. This direct exposure, and access, to the inherent dynamical properties of phononic materials has vigorously chartered a new direction in materials physics, at a multitude of scales, and has already begun to impact numerous applications ranging from vibration control \cite{Phani_2006,*hussein2007dispersive}, through subwavelength sound focusing \cite{yang2004focusing,*zhu2010holey} and cloaking \cite{cummer2007one,*torrent2007acoustic}, to reducing the thermal conductivity of semiconductors \cite{McGaughey_2006,*davis2011thermal,*PhysRevLett.112.055505}. A discussion of applications and references are provided in Refs. \cite{li2012colloquium,Deymier2013,Hussein_AMR_2014}, and recent special journal issues on the topic assemble some of the latest advances in the field \cite{hussein2011preface,*hussein_JVA_2013}.

\subsection{Elastic wave dispersion in the presence of nonlinearity}
The majority of theoretical investigations of wave motion in elastic solids are based on linear analysis, that is, linear constitutive laws and linear strain-displacement relationships are assumed (see Refs.~\cite{graff1975wave} and~\cite{achenbach1984wave}, and references therein). The incorporation of nonlinear effects, however, gives rise to a broader range of physical phenomena including amplitude-dependent wave motion~\cite{bhatnagar1979nonlinear,*ogden1997non,*Norris_1998,porubov2003amplification}. Finite-strain waves in elastic solids is a subset among the broader class of nonlinear waves. From a mathematical perspective, a formal treatment of finite strain requires the incorporation of a nonlinear strain tensor in setting up the governing equations of motion. Regardless of the type of nonlinearity, a common analysis framework has been one in which the dispersion is viewed to arise linearly, e.g., due to the presence of a microstructure or geometrical constraints, and that such dispersion may be balanced with nonlinear effects to allow for the generation of nonlinear traveling waves of fixed spatial profile such as shock waves and solitons~\cite{porubov2003amplification,erofeyev2003wave}. In contrast to this dispersion/non-linearity balancing framework where the focus is on finding these special types of waves and characterizing the amplitude-dependence, or wave-number-dependence, of their speeds, it has recently been shown that nonlinearity in itself may cause dispersion without the need for a linear dispersive mechanism~\cite{abedinnasab2013wave,*Lee_PNAS_2013}. This perspective provides a motivation to derive dispersion relations that inherently embody the effects of the nonlinearities on the dispersion, i.e., amplitude-dependent relations for general wave motion that encompass both the speed (or frequency) and the wave number.  

In the context of nonlinear phononic materials, there are several studies that follow the premise of Bloch wave propagation analysis. These include investigations on systems exhibiting material nonlinearity, analyzed using the method of multiple scales \cite{vakakis1995nonlinear,*manktelow2011multiple,*Swinteck_2013_JVA}, perturbation analysis \cite{chakraborty2001dynamics,*narisetti2010perturbation}, the harmonic balance method \cite{lazarov2007low,*narisetti2012study}, and the transfer matrix (TM) method in conjunction with a perturbation technique~\cite{manktelow2013comparison}. The effects of nonlinearity on the dynamics of periodic materials has also been explored in the context of atomic-scale models incorporating anharmonic potentials; see, for example, a recent study focusing on phonon transport~\cite{Li_2012}. Concerning finite-strain dispersion in a layered elastic medium, this was recently investigated by Andrianov \textit{et al}.~\cite{Andrianov_WM_2013} via a homogenization approach whereby the periodic unit cell was first homogenized as a linear medium and subsequently a finite-strain dispersion relation was derived for the averaged medium. This approach therefore does not retain the periodic character in the derived dispersion relation. On the experimental track, numerous studies have been conducted in recent years on nonlinear wave phenomena particularly in periodic granular chains, e.g.,~\cite{daraio2006tunability,*herbold2009pulse}. It is evident that the effects of nonlinearity in phononic/granular materials could be utilized to enrich the design of devices in numerous engineering applications, such as for shock mitigation~\cite{Daraio_2006_PRL}, tunable wave filtering~\cite{narisetti2010perturbation}, focusing~\cite{Spadoni_2010_PNAS} and rectification~\cite{Boechler_2011_NM}.


\subsection{Overview}
In this paper, we present a theoretical treatment of elastic wave motion in phononic materials in the presence of nonlinearity, specifically the type arising from finite elastic strain. We consider \textit{phononic crystals}, which is a class of phononic materials in which the prime dispersion mechanism is Bragg scattering~\footnote[1]{A phononic material in general may be classified into two types, a phononic crystal and a locally resonant elastic metamaterial~\cite{Deymier2013,Hussein_AMR_2014}. In this work we focus on the former, but the mathematical treatment is also applicable to the latter~\cite{Khajehtourian_Hussein_2013}.}. For ease of exposition, we focus on a one dimensional (1D) layered material model admitting only longitudinal displacements (which may also be viewed as a model for a periodic thin rod). Since the TM method provides the backbone of our approach, we first briefly overview it, in conjunction with Bloch's theorem~\cite{bloch1929quantenmechanik}, for the exact analytical analysis of simple 1D linear phononic crystals (Section \ref{LinearPC}). We then review the treatment of geometric nonlinearity, i.e., finite strain, in the context of a homogeneous medium (Section \ref{NLHomo}). In Section \ref{NLHeter}, we combine the previous derivations, that is, we allow the finite-strain dispersion relation for a homogeneous medium to represent the motion characteristics in a single layer of a periodically layered 1D phononic crystal and subsequently incorporate this relation into the TM formalism. While the finite-strain dispersion within each layer is exact, the dispersion relation we obtain for the overall 1D phononic crystal represents an approximate solution. Finally, we use our formulation to investigate the effects of geometric nonlinearity on the elastic band structure and Bloch mode shapes as a function of the amplitude of motion.

\begin{figure*}
\centering
	\includegraphics[scale=1]{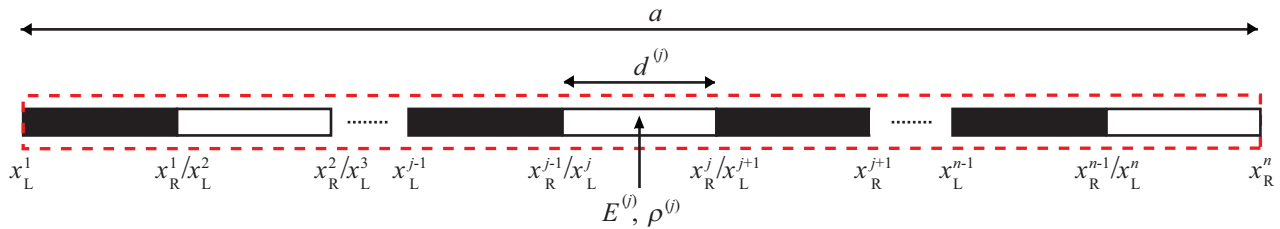}	
		\caption{Continuous model of a 1D two-phased phononic crystal viewed as a periodic thin rod.\label{Fig1}}
\end{figure*}

\section{Wave propagation in 1D linear phononic crystals}\label{LinearPC}
Bloch's theorem~\cite{bloch1929quantenmechanik} provides the underlying mathematical framework for obtaining the elastic band structure (i.e., dispersion curves) for a phononic crystal. There are several approaches for applying the theorem to a unit cell modeled as a continuum. In this work we utilize the TM method, which is described briefly below (for more details, see Hussein \textit{et al}.~\cite{hussein2006dispersive}).

We begin our dynamic analysis of a phononic crystal with the statement of the equation of motion. As mentioned earlier, we restrict ourselves to a 1D model, e.g., a thin rod, for which the equation of motion is
\begin{equation}\label{eq:Euxx}
(Eu_{,x})_{,x}=\rho u_{,tt},
\end{equation}
where $x$, $t$, $u=u(x,t)$, $E=E(x)$ and $\rho=\rho(x)$ denote the position, time, displacement, material Young's modulus and material density, respectively. 

Equation\ (\ref{eq:Euxx}) may be used to study the propagation of elastic waves in various 1D media. In particular, if we have a homogeneous, linearly elastic 1D rod of infinite extent (having no boundaries at which waves may reflect), then we may apply a plane wave solution of the form
\begin{equation}\label{eq:uxt}
u(x,t)=A \mathrm{e}^{\mathrm{i}(\kappa x-\omega t)},
\end{equation}
where $A$ is the wave amplitude, $\kappa$ is the wave number, $\omega$ is the temporal frequency of the traveling wave, and $\mathrm{i}=\sqrt{-1}$. Substituting Eq.\ (\ref{eq:uxt}) into Eq.\ (\ref{eq:Euxx}) provides the linear dispersion relation
\begin{equation}\label{eq:Ek2}
E\kappa^2=\rho\omega^2.
\end{equation}

\noindent This approach may also be applied to heterogeneous media provided the heterogeneity is periodic. In this case, we refer to Eq.\ (\ref{eq:uxt}) as Bloch's theorem (where $A=A(x,\kappa)$), and it suffices to analyze only a single unit cell representing the unique segment that is repeated to generate the periodic medium and to apply periodic boundary conditions to this segment. In Fig. \ref{Fig1}, we present a simple bi-material model of a 1D phononic crystal in the form of a layered periodic rod (where the unit cell is enclosed in a red dashed box). The spatial lattice spacing of the 1D phononic crystal is denoted by the constant $a$. 

For an arbitrary homogeneous layer $j$ in the unit cell, the associated material properties, which are constant, are denoted as $E^{(j)}$ and $\rho^{(j)}$. The longitudinal velocity in layer $j$ is therefore $c^{(j)}=\sqrt{{E^{(j)}}/{\rho^{(j)}}}$. The layer is bordered by layer $j-1$ on the left and layer $j+1$ on the right. Denoting the thickness of an arbitrary layer by $d^{(j)}$, the cell length is $a=\sum_{j=1}^{n}d^{(j)}$ for a unit cell with $n$ layers. Following this notation, the solution to Eq.\ (\ref{eq:Euxx}) is formed from the superposition of forward (transmitted) and backward (reflected) traveling waves with a harmonic time dependence,
\begin{equation}\label{eq:uxtA}
u(x,t)=[A_+^{(j)}\mathrm{e}^{\mathrm{i}\kappa^{(j)}x}+A_-^{(j)}\mathrm{e}^{-\mathrm{i}\kappa^{(j)}x}]\mathrm{e}^{-\mathrm{i}\omega t},
\end{equation}
where $\kappa^{(j)}=\omega/c^{(j)}$ is the layer wave number. The spatial components of Eq. (\ref{eq:uxtA}) may be written along with those of the stress, 
\begin{equation}\label{eq:sigma}
\sigma=Eu_{,x},
\end{equation}
in compact form as
\begin{eqnarray}\label{eq:tmy}
\left[\begin{array}{c} u(x)\\\sigma(x) \end{array}\right]&=&\left[\begin{array}{cc} 1 & 1\\\mathrm{i} Z^{(j)} &-\mathrm{i} Z^{(j)} \end{array}\right]\left[\begin{array}c A_+^{(j)}\mathrm{e}^{\mathrm{i}\kappa^{(j)}x}\\A_-^{(j)}\mathrm{e}^{-\mathrm{i}\kappa^{(j)}x} \end{array}\right]\nonumber\\
&=&\mathbf{H}_j\left[\begin{array}c A_+^{(j)}\mathrm{e}^{\mathrm{i}\kappa^{(j)}x}\\A_-^{(j)}\mathrm{e}^{-\mathrm{i}\kappa^{(j)}x} \end{array}\right],
\end{eqnarray} 
where $Z^{(j)}=E^{(j)}\kappa^{(j)}$. There are two conditions that must be satisfied at the layer interfaces: (1) the continuity of displacement and (2) the continuity of stress. This allows for the substitution of the relation $x_R^{(j)}=x_L^{(j)}+d^{(j)}$ (where $x_R^{(j)}$ and $x_L^{(j)}$ denote the position of the right and left boundary, respectively, of layer $j$) into Eq.\ (\ref{eq:tmy}) and thus relating the displacement and stress at $x_L^{(j)}$ to those at $x_R^{(j)}$. Subsequently, by setting $x=x_L^{(j)}$ in Eq.\ (\ref{eq:tmy}), we get
\begin{eqnarray}
\label{eq:tmT}
\left[\begin{array}{c} u(x_R^{(j)})\\\sigma (x_R^{(j)}) \end{array}\right]\!\!=\! \mathbf{H}_j \mathbf{D}_j \mathbf{H}_j^{-1} \! \left[\begin{array}{c} u(x_L^{(j)})\\\sigma (x_L^{(j)}) \end{array}\right]\!\!=\!\mathbf{T}_j \! \left[\begin{array}{c} u(x_L^{(j)})\\\sigma (x_L^{(j)}) \end{array}\right]\!\!,
\end{eqnarray}

\noindent where
\begin{eqnarray}
\mathbf{D}_j \! = \left[\begin{array}{cc} \mathrm{e}^{\mathrm{i}\kappa^{(j)}d^{(j)}} & 0  \\ 0 & \mathrm{e}^{-\mathrm{i}\kappa^{(j)}d^{(j)}} \end{array}\right] \! \!, 
\end{eqnarray}

\noindent and $\mathbf{T}_j$, the \textit{transfer matrix} for layer $j$, has the expanded form
\begin{eqnarray}\label{eq:tm}
\mathbf{T}_j\!=\!\!\left[\!\!\begin{array}{cc}   \cos{(\kappa^{(j)}d^{(j)})}    &    ({1}/{Z^{(j)}})\sin{(\kappa^{(j)}d^{(j)})} \\[6pt] -Z^{(j)}\sin{(\kappa^{(j)}d^{(j)})} &    \cos{(\kappa^{(j)}d^{(j)})} \end{array}\!\!\right]\!\!.
\end{eqnarray} 

As previously stated, Eq.\ (\ref{eq:tmT}) relates the displacement and stress at $x_L^{(j)}$ to those at $x_R^{(j)}$ of the same layer $j$. However, since the construction of the transfer matrix is valid for any layer and $x_L^{(j)}\equiv x_R^{(j-1)}$, the result in Eq.\ (\ref{eq:tmT}) can be extended recursively across several layers. In the interest of brevity, let $\bf{y}(.)=\rm [u(.)\:\sigma(.)]^{\rm T}$, thus,
\begin{equation}\label{eq:yxR}
\mathbf{y}(x_R^{n})=\mathbf{T}_n\mathbf{T}_{n-1}\hdots\mathbf{T}_1\mathbf{y}(x_L^{1})=\mathbf{T}\mathbf{y}(x_L^{1}).
\end{equation}
Ultimately, the displacement and stress at the left end of the first layer ($x=x_L^{1}$) in a unit cell are related to those at the right boundary of the $n$th layer ($x=x_R^{n}$) by the cumulative transfer matrix, $\mathbf{T}$.

Now we turn to Bloch's theorem, which states that the time harmonic response at a given point in a unit cell is the same as that of the corresponding point in an adjacent unit cell except for a phase difference of $\mathrm{e}^{\mathrm{i}\kappa a}$. This relation is given by $f(x+a)=\mathrm{e}^{\mathrm{i}\kappa a}f(x)$, which when applied to the states of displacement and stress across a unit cell gives
\begin{eqnarray}\label{eq:yxnR}
\mathbf{y}(x_R^{n})=\mathrm{e}^{\mathrm{i}\kappa a} \mathbf{y}(x_L^{1}).
\end{eqnarray} 

Combining Eqs.\ (\ref{eq:yxR}) and (\ref{eq:yxnR}) yields the eigenvalue problem
\begin{eqnarray}\label{eq:TI}
[\mathbf{T}-\mathbf{I}\gamma]\mathbf{y}(x_L^{1})=\mathbf{0},
\end{eqnarray} 
where $\gamma=\mathrm{e}^{\mathrm{i}\kappa a}$. The solution of Eq.\ (\ref{eq:TI}), which appears in complex conjugate pairs, provides the dispersion relation $\kappa=\kappa(\omega)$ for the 1D phononic crystal. Real-valued wave numbers, calculated from $\gamma$ using Eq.\ (\ref{eq:Re}), support propagating wave modes, whereas imaginary wave numbers, extracted from $\gamma$ using Eq.\ (\ref{eq:Im}), represent spatially attenuating modes:
\begin{eqnarray}\label{eq:ReIm}
\kappa_\mathrm{R}=\frac{1}{a}\mathrm{Re}[\frac{1}{\mathrm{i}}\mathrm{ln}\gamma],\label{eq:Re}\\ [6pt]
\kappa_\mathrm{I}=\frac{1}{a}\mathrm{Im}[\frac{1}{\mathrm{i}}\mathrm{ln}\gamma].\label{eq:Im}
\end{eqnarray} 

\section{Treatment of nonlinearity}\label{NonLinear}

We now provide a theoretical treatment of finite-strain dispersion; first we review the prerequisite problem of a 1D homogeneous medium, and follow with the derivation for a 1D phononic crystal. In the homogeneous medium problem, the approach is exact regardless of the amplitude of the traveling wave. In the subsequent derivation of the phononic crystal dispersion curves, the accuracy decreases with increasing wave amplitude. 

\subsection{Finite-strain waves in 1D homogeneous media}\label{NLHomo}
The equation of motion and finite-strain dispersion relation is reviewed here for 1D plane wave motion in a bulk homogeneous medium without consideration of lateral effects. In principle, this problem is equivalent to that of a slender rod. In the derivations, all terms in the nonlinear strain tensor are retained and no high order terms emerging from the differentiations are subsequently neglected. The reader is referred to Ref. \cite{abedinnasab2013wave} for more details as well as a validation of the theoretical approach by means of a comparison with a standard finite-strain numerical simulation of a corresponding 1D model with finite dimensions. 

\subsubsection{Equation of motion}
Introducing $u$ as the elastic longitudinal displacement, the exact complete Green-Lagrange strain field in our 1D model is given by
\begin{equation}\label{eq:eps}
\epsilon=\frac{\partial u}{\partial s}+\frac{1}{2}(\frac{\partial u}{\partial s})^2,
\end{equation}
where the first and second terms on the RHS represent the linear and nonlinear parts, respectively, and $s$ is the Lagrangian longitudinal coordinate which is equal to $x$ in Eq.\ (\ref{eq:Euxx}).

Using Hamilton's principle, we write the equation of motion under longitudinal stress as
\begin{eqnarray}\label{eq:var}
	\int_{0}^{t}(\delta T-\delta U^e)\mathrm{d}t=0,
\end{eqnarray}
where $T$ and $U^e$ denote kinetic and elastic potential energies, respectively. We note that no external nonconservative forces and moments are permitted because of our interest in the free wave propagation problem. Furthermore, the effects of lateral inertia are neglected. The variation of kinetic energy is obtained using integration by parts and is given as
\begin{eqnarray}\label{eq:dT}
	\delta T=-\rho A\int_{0}^{l}(u_{,tt}\delta u)\mathrm{d}s,
\end{eqnarray}
\noindent where $l$ denotes the length of a portion of the 1D medium. Similarly, the variation of elastic potential energy is written as
\begin{eqnarray}\label{eq:dU}
	\delta U^e=\int_{0}^{l}\int_{A}(\sigma\delta\epsilon)\mathrm{d}A\,\mathrm{d}s,
\end{eqnarray}
where $\sigma$ is the longitudinal stress. We choose to base our analysis on the Cauchy stress and model the stress-strain relationship by Hooke's law, $\sigma=E\epsilon$. Using Eq.\ (\ref{eq:dU}), and with the aid of integration by parts, we can now write the variation of elastic potential energy as
\begin{eqnarray}\label{eq:dUe}
	\delta U^e=\int_{0}^{l}\{\frac12EAh(h^2-1)\delta u^\prime\}\mathrm{d}s,
\end{eqnarray}
where $u^\prime=du/ds=u_{,s}$, and $h$ is an agent variable defined as
\begin{eqnarray}\label{eq:h}
	h=1+u^\prime.
\end{eqnarray}
Substitution of  Eqs.\ (\ref{eq:dUe}) and (\ref{eq:dT}) into Eq.\ (\ref{eq:var}) produces the exact finite-strain equation of motion as
\begin{eqnarray}\label{eq:A1}
\rho Au_{,tt}=\frac12EA(3h^2-1)u^{\prime\prime}.
\end{eqnarray}
If the longitudinal deformation is infinitesimal, then $u^\prime$ is small and from Eq.\ (\ref{eq:h}), $h\approx 1$. Substitution of $h=1$ into Eq.\ (\ref{eq:A1}) leads to
\begin{eqnarray}\label{eq:A11}
\rho Au_{,tt}=EAu^{\prime\prime},
\end{eqnarray}
which is the equation of motion describing infinitesimal longitudinal deformation.

\subsubsection{Dispersion relation}
Using Eq.\ (\ref{eq:h}), we rewrite Eq.\ (\ref{eq:A1}) as 
\begin{eqnarray}\label{eq:utt}
	u_{,tt}-c^2u^{\prime\prime}=\frac12\Big[c^2[3(u^\prime)^{2}+(u^\prime)^3]\Big]^\prime,
\end{eqnarray}
where $c=\sqrt{E/\rho}$. Differentiation of Eq.\ (\ref{eq:utt}) with respect to $s$ gives
\begin{eqnarray}\label{eq:uttprime}
	(u_{,tt})^\prime-c^2u^{(3)}=\frac12\Big[c^2[3(u^\prime)^2+(u^\prime)^3]\Big]^{\prime\prime}.
\end{eqnarray}
Defining $\bar{u}=u^\prime$ and $z=\lvert\kappa\lvert s+\omega_{\mathrm{fin}}t$, where $\omega_\mathrm{fin}$ represents the wave frequency under finite strain, Eq.\ (\ref{eq:uttprime}) becomes
\begin{eqnarray}\label{eq:uzz}
	\omega_{\mathrm{fin}}^2\bar{u}_{,zz}-c^2\kappa^2\bar{u}_{,zz}=\frac12\kappa^2\Big[c^2[3\bar{u}^2+\bar{u}^3]\Big]_{,zz}.
\end{eqnarray}
Integrating Eq.\ (\ref{eq:uzz}) twice leads to
\begin{eqnarray}\label{eq:wfin2}
	(\omega_{\mathrm{fin}}^2-c^2\kappa^2)\bar{u}-\frac{c^2\kappa^2}{2}[3\bar{u}^2+\bar{u}^3]=0,
\end{eqnarray}
where the nonzero constants of integration (in the form of polynomials in $z$) represent secular terms which we have set equal to zero in light of our interest in the dispersion relation. Selecting the positive root of Eq.\ (\ref{eq:wfin2}) we get
\begin{eqnarray}\label{eq:ubarz}
\bar{u}(z)=\frac{-3+\sqrt{1+8\omega_{\mathrm{fin}}^2/c^2\kappa^2}}{2}.
\end{eqnarray}
Since $\bar{u}=u_{,s}$, we recognize that $\bar{u}=\lvert\kappa\lvert u_{,z}$ and therefore Eq.\ (\ref{eq:ubarz}) represents a first-order ordinary differential equation with $z$ and $u$ as the independent and dependent variables, respectively.

\begin{figure}
\centering
	\includegraphics[scale=1]{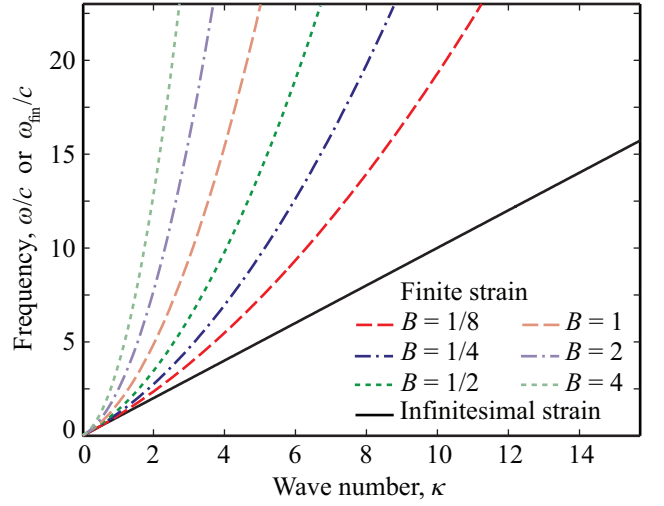}	
		\caption{Frequency dispersion curves for a 1D homogeneous elastic medium~\cite{abedinnasab2013wave}. The finite-strain dispersion relation is based on Eq.\ (\ref{eq:wfin}); the infinitesimal strain dispersion relation is based on Eq.\ (\ref{eq:wk}).\label{Fig2}}
\end{figure}

Now we return to Eq.\ (\ref{eq:utt}) and consider for initial conditions a sinusoidal displacement field, with amplitude $B$ and a zero phase in time, and a zero velocity field. This represents a fundamental harmonic signal for which we seek to characterize its dispersive behavior. In principle, any choice of the initial velocity field is permitted. Following the change of variables that has been introduced, these initial conditions correspond to the following restrictions at $z=0$ on the $\bar{u}(z)$ function given in Eq.\ (\ref{eq:ubarz}):
\begin{eqnarray}\label{eq:ubarzero}
\bar{u}(0)=\lvert\kappa\lvert B, \;\; \bar{u}_{,z}(0)=0.
\end{eqnarray}
These represent initial conditions in the wave phase, $z$, for Eq.\ (\ref{eq:uzz}) and allow for the introduction of the wave amplitude, $B$, into the formulation. Applying Eq.\ (\ref{eq:ubarzero}) to Eq.\ (\ref{eq:ubarz}) enables us to use the latter to solve for $\omega_{\mathrm{fin}}$ for a given value of $\kappa$. This leads to the exact dispersion relation,
\begin{eqnarray}\label{eq:wfin}	\omega_{\mathrm{fin}}(\kappa;B)=\sqrt{\frac{2+3B\lvert\kappa\lvert+(B\kappa)^2}{2}}\omega,
\end{eqnarray}
where $\omega$ is the frequency based on infinitesimal strain,
\begin{eqnarray}\label{eq:wk}
	\omega(\kappa)=c\lvert\kappa\lvert.
\end{eqnarray}
By taking the limit, $\mathrm{lim}_{B\rightarrow 0}\,\omega_{\mathrm{fin}}(\kappa;B)$, in Eq.\ (\ref{eq:wfin}) we recover Eq.\ (\ref{eq:wk}) which is the standard linear dispersion relation for a 1D homogeneous elastic medium or a thin rod~\cite{billingham2000wave}.

For demonstration, six amplitude-dependent finite-strain dispersion curves based on Eq.\ (\ref{eq:wfin}) are plotted in Fig.\ \ref{Fig2}. These curves describe the fundamental dispersive properties that emerge due to the incorporation of finite strain. The curves demonstrate that nonlinearity by itself causes wave dispersion in an elastic medium, i.e., without the need for a linear dispersive mechanism. From a physical point of view one may envision an initial prescribed harmonic wave being set free at some point in time. The dispersion relation of Eq.\ (\ref{eq:wfin}) describes the frequency versus wave number relation for this wave as it disperses in the presence of amplitude-dependent finite strain. This concept was tested numerically and validated in Ref. \cite{abedinnasab2013wave}. Superimposed in  Fig.\ \ref{Fig2} is the dispersion curve based on infinitesimal strain, i.e., Eq.\ (\ref{eq:wk}). It is noted that the deviation between a finite-strain curve and the infinitesimal-strain curve increases with wave number, and the effect of the wave amplitude on this deviation is illustrated by an accumulative doubling of the amplitude across the six finite-strain cases shown. 

\subsection{Finite-strain waves in 1D phononic crystals}\label{NLHeter}
The TM method is now used to obtain a dispersion relation for a 1D phononic crystal whose constituent materials are exhibiting finite-strain dispersion. The outcome is an approximate overall dispersion relation since the construction of the transfer matrix is based on a linear strain-displacement relationship [see Eq. (\ref{eq:sigma}) and (\ref{eq:tmy})]. While not exact, this approach provides a quantitative prediction of the effects of nonlinearity on the location and size of band gaps and the values of the group velocity across the spectrum, all as a function of wave amplitude. While the technique is not limited to small values of $B/a$, its accuracy reduces as the strength of the nonlinearity increases.\\ 
\begin{figure*}
	\includegraphics[scale=1]{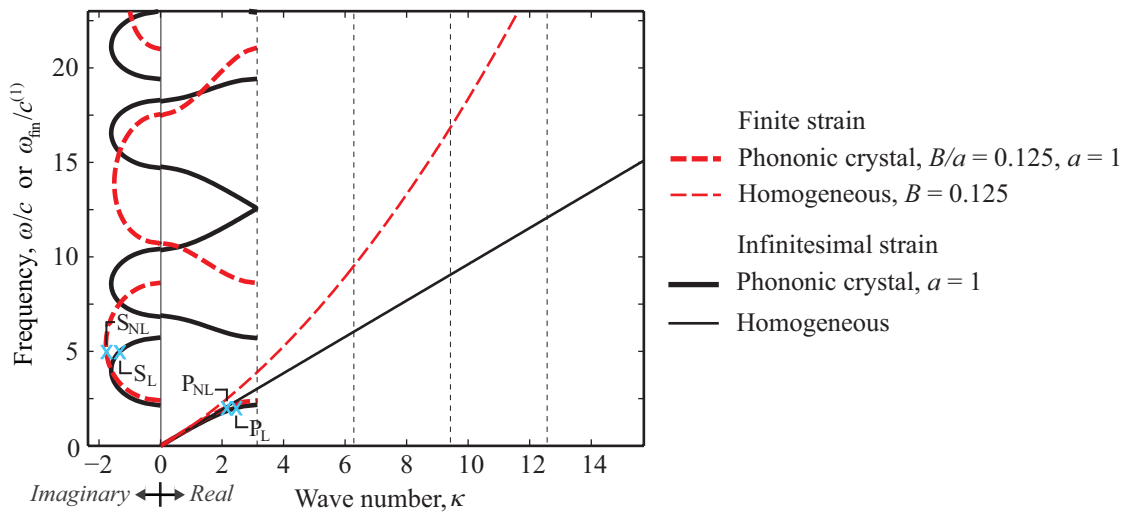}	
		\caption{Frequency band structure for a 1D phononic crystal under finite strain [obtained using Eq.\ (\ref{eq:roots})]. For comparison, the dispersion curves under infinitesimal strain are included. Also, corresponding dispersion curves for a statically equivalent 1D homogeneous elastic medium are overlaid. The nonlinearity-induced shifting of the dispersion curves is marked at two frequencies. Points $\rm P_{\rm L}$ and $\rm P_{\rm NL}$ are at frequency $\omega/c^{(1)}= 2$ and lie on the first infinitesimal-strain and the first finite-strain pass-band branch, respectively. Points $\rm S_{\rm L}$ and $\rm S_{\rm NL}$ are at frequency $\omega/c^{(1)}= 5$ and lie on the first infinitesimal-strain and the first finite-strain stop-band branch, respectively. \label{Fig3}}
\end{figure*}
\begin{figure*}
 \includegraphics{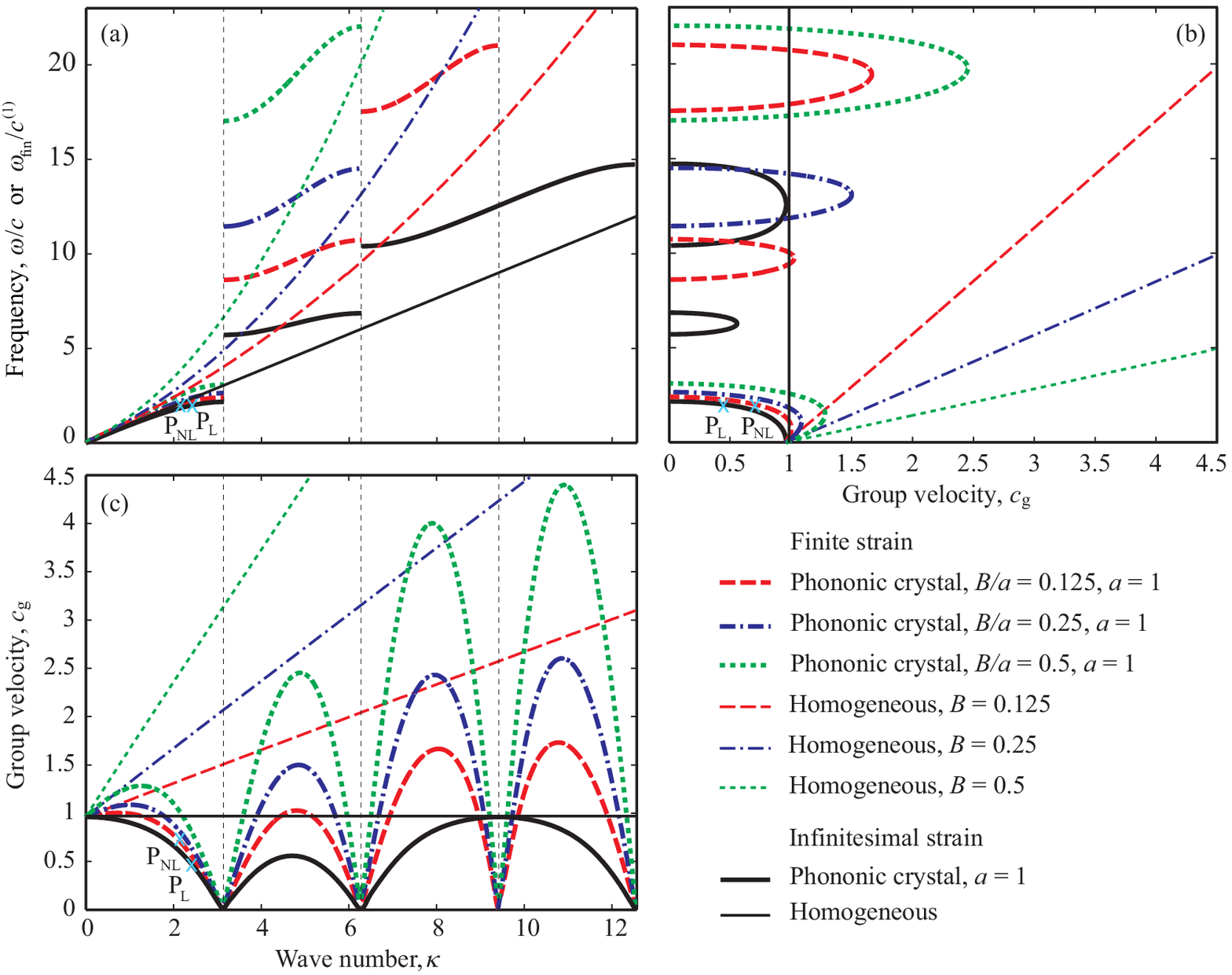}
  \caption{Effect of nonlinearity on group velocity for the 1D phononic crystal and the statically equivalent 1D homogeneous elastic medium considered in Fig.~\ref{Fig3}. (a) unfolded frequency band structure, (b) frequency versus group velocity, (c) group velocity versus wave number. The nonlinearity-induced shifting of the dispersion curves at frequency $\omega /c^{(1)}= 2$ is noted. \label{Fig4} }
\end{figure*}
As presented in Section \ref{LinearPC}, the TM method is applicable in either the absence or presence of nonlinearity; the distinction is made in
the definition of $\kappa^{(j)}$ in Eq.~(\ref{eq:uxtA}). For the linear problem, $\kappa^{(j)}=\omega/c^{(j)}$ as outlined earlier. A similar relationship between the wave number and the finite-strain wave frequency, $\omega_{\rm fin}$, may be developed.

First we rewrite Eq.\ (\ref{eq:wfin}) explicitly for layer $j$,
\begin{eqnarray}\label{eq:wfinpc}
	\omega_{\mathrm{fin}}=c^{(j)}\kappa^{(j)}\sqrt{\frac{2+3B\kappa^{(j)}+[B\kappa^{(j)}]^2}{2}},
\end{eqnarray}
which may be cast as the following 4th order characteristic equation:
\begin{eqnarray}\label{eq:4th}
 [\kappa^{(j)}]^2(1+B\kappa^{(j)})(2+B\kappa^{(j)})-2\frac{\omega_{\mathrm{fin}}^2}{[c^{(j)}]^2}=0.
\end{eqnarray}
Solving Eq.\ (\ref{eq:4th}) gives
\begin{subequations}\label{eq:roots}
\begin{eqnarray}
  \kappa_{1,2}^{(j)}&=&\frac{1}{12 B}\Big(-9+P^{(j)}\mp\sqrt{Q^{(j)}-R^{(j)}}\Big), \label{eq:roots1} \\[6pt]
  \kappa_{3,4}^{(j)}&=&-\frac{1}{12 B}\Big(9+P^{(j)}\pm\sqrt{Q^{(j)}+R^{(j)}}\Big),\label{eq:roots2}
 \end{eqnarray}
\end{subequations}
where
\begin{subequations}\label{eq:PQRS}

\begin{eqnarray}
P^{(j)}\!\!&=&\!\!\sqrt{\!\frac{33c^{(j)}A^{(j)}\!+\!12(4[c^{(j)}]^2\!-\!24B^2\omega_{\mathrm{fin}}^2\!+\![A^{(j)}]^2)}{c^{(j)}A^{(j)}}},\label{eq:P}\,\,\,\,\,\,\,\,\,\,\,\,\,\\[3pt]
Q^{(j)}\!\!&=&\!\!\frac{66 c^{(j)} A^{(j)}\!-\!48([c^{(j)}]^2-6B^2 \omega_{\mathrm{fin}}^2)\!-\!12[A^{(j)}]^2}{c^{(j)}A^{(j)}},\label{eq:Q}\,\,\,\,\,\,\,\,\,\,\,\,\\[3pt]
R^{(j)}\!\!&=&\!\!\frac{54\sqrt{3c^{(j)} A^{(j)}}} {\sqrt{{\!11 c_0^{(j)}\! A^{(j)}\!+\!4(4[c^{(j)}]^2\!+\![A^{(j)}]^2\!-24 B^2 \omega_{\mathrm{fin}}^2 )}}},\,\,\,\,\,\,\,\,\,\,\,\,\label{eq:R}
  \end{eqnarray}
 and
 \begin{eqnarray}\label{eq:S}
A^{(j)}\!\!&=&\!\!\Bigg(\!\!\!-\!99B^2 \omega_{\mathrm{fin}}^2c^{(j)}\!+\!8[c^{(j)}]^3\nonumber\\
+\!3B&\omega_{\mathrm{fin}}&\sqrt{\!(1536B^2\!+\!321[c^{(j)}]^2)B^2\omega_{\mathrm{fin}}^4\!\!-\!\!48[c^{(j)}]^4}\Bigg)^{\!\!\!\frac{1}{3}}\!\!.\,\,\,\,\,\,\,\,\,\,\,\,
 \end{eqnarray}
\end{subequations}

Now we will use Eq.\ (\ref{eq:roots}) and the TM method to obtain an approximation of the finite-strain dispersion curves of a 1D phononic crystal that has the same geometric features as the periodic bi-material rod in Fig. \ref{Fig1} and the following ratio of material properties: $c^{(2)}/c^{(1)}=2$ and $\rho^{(2)}/\rho^{(1)}=3$. We consider a bi-layered unit cell in which $d^{(2)}=d^{(1)}$. The results are shown in Fig. \ref{Fig3} for a phononic crystal of size $a=1$ (arbitrary units) and a value of wave amplitude of $B/a=0.125$. Superimposed, for comparison, are the dispersion curves on the basis of infinitesimal strain and the corresponding dispersion curves for an equivalent statically homogenized medium for which the speed of sound is $c$ (obtained by the standard rule of mixtures). We observe in the figure that the finite-strain dispersion curves asymptotically converge to the infinitesimal and homogenized curves at long wavelengths as expected. We also note that the finite strain causes the dispersion branches to rise and the band-gap sizes to increase significantly$-$an attractive trait for many applications involving sound and vibration control. This behavior, however, is dependent on the type of nonlinearity considered.\\
\begin{figure*}
	\includegraphics{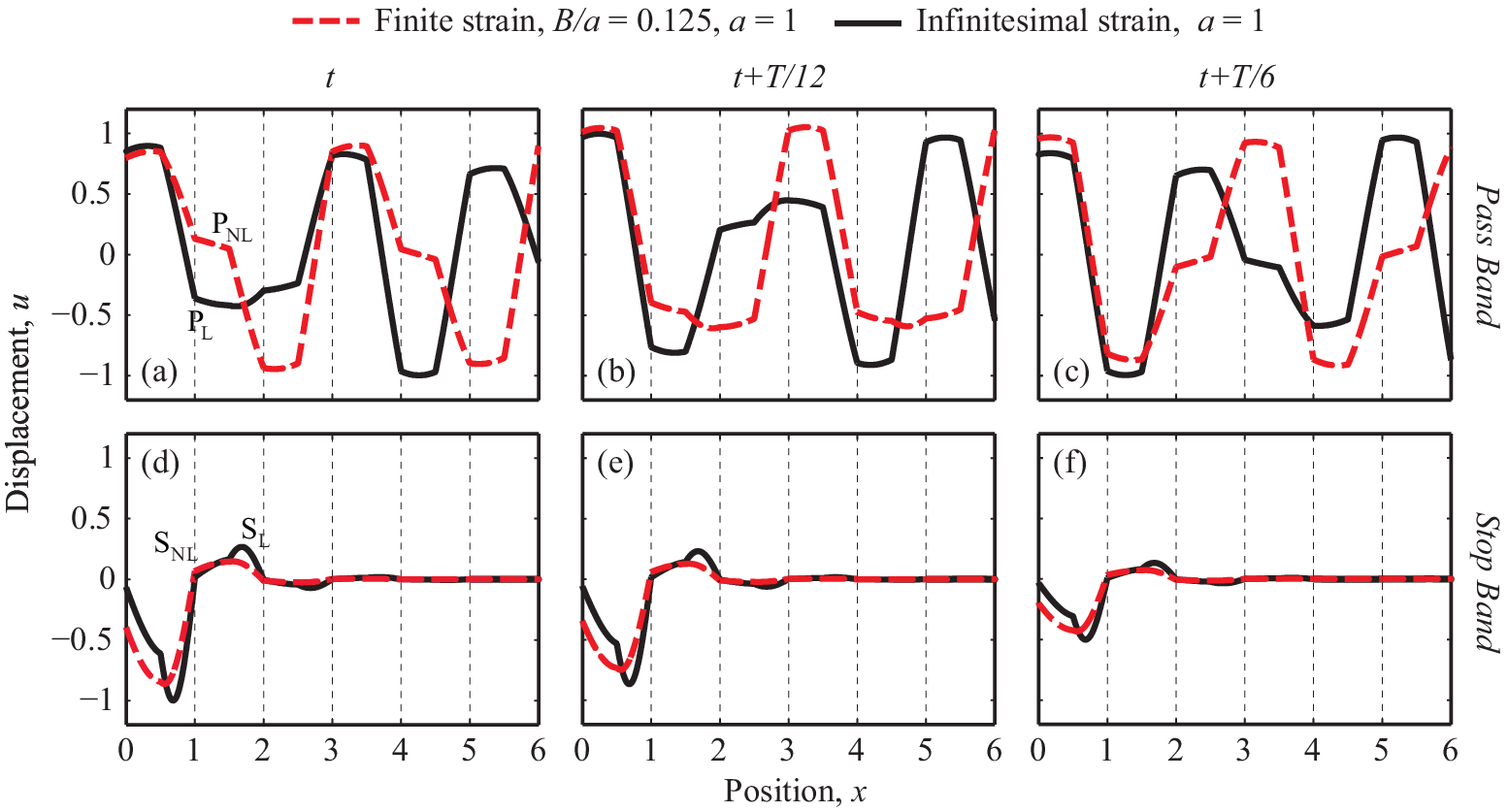}
		\caption{Three time snap shots of the Bloch mode shape over six unit cells corresponding to the four points $\rm P_{\rm L}$, $\rm P_{\rm NL}$, $\rm S_{\rm L}$, and $\rm S_{\rm NL}$ marked in Fig.~\ref{Fig3}. The Bloch mode shape for points $\rm P_{\rm L}$ and $\rm P_{\rm NL}$ are shown in (a), (b), and (c), while the Bloch mode shapes for points $\rm S_{\rm L}$ and $\rm S_{\rm NL}$ are shown in (d), (e), and (f). Red dashed curves correspond to finite strain and black solid curves correspond to infinitesimal strain. Each pass-band and stop-band sets of curves are normalized with respect to the maximum displacement value of the infinitesimal strain case at time $t$.\label{Fig5}}
\end{figure*}  
The influence of the nonlinearity on the frequency-wave number relation naturally impacts the spectrum of group velocities, defined as
\begin{eqnarray}\label{eq:GrVel}
 c_\mathrm{g}=\frac{\partial \omega_{\mathrm{fin}}(\kappa;B)}{\partial \kappa}.
\end{eqnarray}
In Fig. \ref{Fig4}, we show the amplitude-dependent relationship between the frequency and the group velocity and between the group velocity and the wave number. The unfolded frequency band structure is also included for correlation. Most noticeable in this figure is the significant rise in the group velocity with amplitude. A similar rise takes effect for the phase velocity as well (not shown), indicating that with finite strain, the medium's permissible wave speeds are supersonic with respect to the nominal speeds under linear, infinitesimal strain. We also note that the homogenized medium's group velocity curves under finite strain are linear and exceed the maximum group velocity values for the corresponding phononic crystal; whereas, in contrast, the maximum group velocity in the infinitesimal-strain problem overlaps with the corresponding homogenized medium's horizontal group velocity line. This disparity may be a manifestation of the linear approximation inherent in the TM method. Thus the minimum distance between the maximum group velocity of a phononic crystal and the corresponding homogenized medium's group velocity line may be viewed as a measure of accuracy for a given value of wave amplitude $B/a$.

In Fig.~\ref{Fig5}, we show three time snap shots of Bloch mode shapes corresponding to the pair of isofrequency pass-band points (top row) and the pair of isofrequency stop-band points (bottom row) marked in Fig.~\ref{Fig3}. The increase in the wavelength due to the nonlinearity at a given frequency, e.g., by comparing point $\rm P_{\rm NL}$ to $\rm P_{\rm L}$, is observed in Fig.~\ref{Fig5} in the form of a slight stretching of the waveform. The effect of the nonlinearity on the group velocity, as indicated in Fig.~\ref{Fig4}, is less obvious in the mode shape diagrams. The effect of the nonlinearity on stop-band stationery waves is shown to be a strengthening of the spatial attenuation, as predicted from Fig.~\ref{Fig3} at the selected frequency. 
\begin{figure*}
	\includegraphics{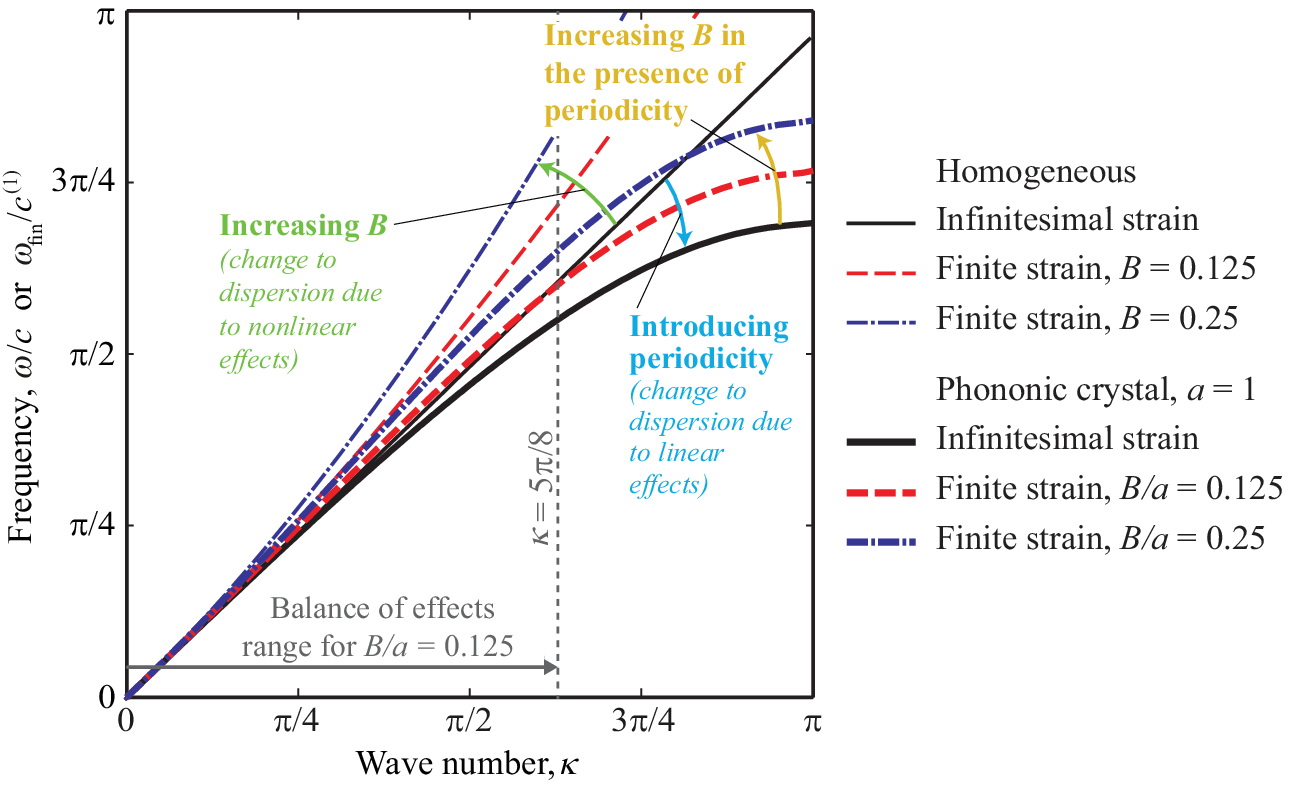}
		\caption{Illustration of the contrast between the effect of dispersion brought about by the introduction of periodicity versus the effect of nonlinearity brought about by increasing the wave amplitude while accounting for finite-strain motion. In the nonlinear periodic medium considered, the two opposing effects are simultaneously present and a balance may be practically realized up to a certain wave number. For a wave amplitude of $B/a=0.125$, the two effects are approximately in balance up to $\kappa=5\pi/8$. \label{Fig6}}
\end{figure*} 


In the literature, a common view is that nonlinearity of the type considered in this study tends to steepen, and subsequently narrow, a wave because large-amplitude constituent waves are able to catch up with slower low-amplitude ones; and, in contrast, dispersion causes a wave to widen its profile spatially because different constituent waves travel at different speeds~\cite{nayfeh2008nonlinear}. Here, we view this problem from a different perspective. We consider the effect of the periodicity in altering the dispersion (which is a linear mechanism) and, in parallel, the effect of nonlinearity on also altering the dispersion. In Fig.~\ref{Fig6}, we reproduce the results displayed in Fig.~\ref{Fig3} with a focus on the first Brillouin zone and with the addition of curves corresponding to a wave amplitude of $B/a=0.25$. The figure illustrates the two effects when taking place separately or in combination. In the structure considered, we observe that an amplitude of $B/a=0.125$ allows the two effects to be practically in balance up to approximately $\kappa=5\pi/8$, which roughly corresponds to a wavelength as small as three times the unit cell size$-$such condition, in principle, may bring rise to a solitary-type wave.

\section{Conclusions}\label{Concl} 
We have theoretically derived a wave number versus frequency finite-strain dispersion relation for Bloch wave propagation in a 1D phononic crystal (layered periodic elastic medium). The effect of finite strain has been incorporated exactly at the individual homogeneous layer level. Subsequently, the TM method has been applied to the unit cell to analytically provide an approximate dispersion relation for the periodic medium. Due to the assumption of a linear strain-displacement gradient relation in the TM method, this approach becomes less accurate as the strength of the nonlinearity increases.   

The results provide a quantitative prediction of the changes in the dispersion curves when periodicity and finite strain are introduced separately or in combination. In particular, we have shown that the wave amplitude could be chosen to create an approximate balance between the two effects up to a certain wave number, as illustrated in Fig.~\ref{Fig6}.    
   
The dynamic behavior revealed by Eqs.\ (\ref{eq:wfin}) and (\ref{eq:roots}) is based on our underlying assumption of Green-Lagrange strain at the homogeneous layer level. Other strain measures could in principle lead to qualitatively different dispersion behavior, and thus it is necessary for future work to examine the problem experimentally to determine the most appropriate strain measure guided by the theory presented in this paper.

\section*{Acknowledgements}
This work has been supported by the National Science Foundation through its CAREER Grant No. 1254931.

\bibliography{PREBibTex}

\begin{thebibliography}{44}%
\makeatletter
\providecommand \@ifxundefined [1]{%
 \@ifx{#1\undefined}
}%
\providecommand \@ifnum [1]{%
 \ifnum #1\expandafter \@firstoftwo
 \else \expandafter \@secondoftwo
 \fi
}%
\providecommand \@ifx [1]{%
 \ifx #1\expandafter \@firstoftwo
 \else \expandafter \@secondoftwo
 \fi
}%
\providecommand \natexlab [1]{#1}%
\providecommand \enquote  [1]{``#1''}%
\providecommand \bibnamefont  [1]{#1}%
\providecommand \bibfnamefont [1]{#1}%
\providecommand \citenamefont [1]{#1}%
\providecommand \href@noop [0]{\@secondoftwo}%
\providecommand \href [0]{\begingroup \@sanitize@url \@href}%
\providecommand \@href[1]{\@@startlink{#1}\@@href}%
\providecommand \@@href[1]{\endgroup#1\@@endlink}%
\providecommand \@sanitize@url [0]{\catcode `\\12\catcode `\$12\catcode
  `\&12\catcode `\#12\catcode `\^12\catcode `\_12\catcode `\%12\relax}%
\providecommand \@@startlink[1]{}%
\providecommand \@@endlink[0]{}%
\providecommand \url  [0]{\begingroup\@sanitize@url \@url }%
\providecommand \@url [1]{\endgroup\@href {#1}{\urlprefix }}%
\providecommand \urlprefix  [0]{URL }%
\providecommand \Eprint [0]{\href }%
\providecommand \doibase [0]{http://dx.doi.org/}%
\providecommand \selectlanguage [0]{\@gobble}%
\providecommand \bibinfo  [0]{\@secondoftwo}%
\providecommand \bibfield  [0]{\@secondoftwo}%
\providecommand \translation [1]{[#1]}%
\providecommand \BibitemOpen [0]{}%
\providecommand \bibitemStop [0]{}%
\providecommand \bibitemNoStop [0]{.\EOS\space}%
\providecommand \EOS [0]{\spacefactor3000\relax}%
\providecommand \BibitemShut  [1]{\csname bibitem#1\endcsname}%
\let\auto@bib@innerbib\@empty
\bibitem [{\citenamefont {Phani}\ \emph {et~al.}(2006)\citenamefont {Phani},
  \citenamefont {Woodhouse},\ and\ \citenamefont {Fleck}}]{Phani_2006}%
  \BibitemOpen
  \bibfield  {author} {\bibinfo {author} {\bibfnamefont {A.~S.}\ \bibnamefont
  {Phani}}, \bibinfo {author} {\bibfnamefont {J.}~\bibnamefont {Woodhouse}}, \
  and\ \bibinfo {author} {\bibfnamefont {N.~A.}\ \bibnamefont {Fleck}},\
  }\href@noop {} {\bibfield  {journal} {\bibinfo  {journal} {J. Acoust. Soc.
  Am.}\ }\textbf {\bibinfo {volume} {119}},\ \bibinfo {pages} {1995} (\bibinfo
  {year} {2006})}\BibitemShut {NoStop}%
\bibitem [{\citenamefont {Hussein}\ \emph {et~al.}(2007)\citenamefont
  {Hussein}, \citenamefont {Hulbert},\ and\ \citenamefont
  {Scott}}]{hussein2007dispersive}%
  \BibitemOpen
  \bibfield  {author} {\bibinfo {author} {\bibfnamefont {M.~I.}\ \bibnamefont
  {Hussein}}, \bibinfo {author} {\bibfnamefont {G.~M.}\ \bibnamefont
  {Hulbert}}, \ and\ \bibinfo {author} {\bibfnamefont {R.~A.}\ \bibnamefont
  {Scott}},\ }\href@noop {} {\bibfield  {journal} {\bibinfo  {journal} {J.
  Sound Vib.}\ }\textbf {\bibinfo {volume} {307}},\ \bibinfo {pages} {865}
  (\bibinfo {year} {2007})}\BibitemShut {NoStop}%
\bibitem [{\citenamefont {Yang}\ \emph {et~al.}(2004)\citenamefont {Yang},
  \citenamefont {Page}, \citenamefont {Liu}, \citenamefont {Cowan},
  \citenamefont {Chan},\ and\ \citenamefont {Sheng}}]{yang2004focusing}%
  \BibitemOpen
  \bibfield  {author} {\bibinfo {author} {\bibfnamefont {S.}~\bibnamefont
  {Yang}}, \bibinfo {author} {\bibfnamefont {J.~H.}\ \bibnamefont {Page}},
  \bibinfo {author} {\bibfnamefont {Z.}~\bibnamefont {Liu}}, \bibinfo {author}
  {\bibfnamefont {M.~L.}\ \bibnamefont {Cowan}}, \bibinfo {author}
  {\bibfnamefont {C.~T.}\ \bibnamefont {Chan}}, \ and\ \bibinfo {author}
  {\bibfnamefont {P.}~\bibnamefont {Sheng}},\ }\href@noop {} {\bibfield
  {journal} {\bibinfo  {journal} {Phys. Rev. Lett.}\ }\textbf {\bibinfo
  {volume} {93}},\ \bibinfo {pages} {024301} (\bibinfo {year}
  {2004})}\BibitemShut {NoStop}%
\bibitem [{\citenamefont {Zhu}\ \emph {et~al.}(2010)\citenamefont {Zhu},
  \citenamefont {Christensen}, \citenamefont {Jung}, \citenamefont
  {Martin-Moreno}, \citenamefont {Yin}, \citenamefont {Fok}, \citenamefont
  {Zhang},\ and\ \citenamefont {Garcia-Vidal}}]{zhu2010holey}%
  \BibitemOpen
  \bibfield  {author} {\bibinfo {author} {\bibfnamefont {J.}~\bibnamefont
  {Zhu}}, \bibinfo {author} {\bibfnamefont {J.}~\bibnamefont {Christensen}},
  \bibinfo {author} {\bibfnamefont {J.}~\bibnamefont {Jung}}, \bibinfo {author}
  {\bibfnamefont {L.}~\bibnamefont {Martin-Moreno}}, \bibinfo {author}
  {\bibfnamefont {X.}~\bibnamefont {Yin}}, \bibinfo {author} {\bibfnamefont
  {L.}~\bibnamefont {Fok}}, \bibinfo {author} {\bibfnamefont {X.}~\bibnamefont
  {Zhang}}, \ and\ \bibinfo {author} {\bibfnamefont {F.}~\bibnamefont
  {Garcia-Vidal}},\ }\href@noop {} {\bibfield  {journal} {\bibinfo  {journal}
  {Nat. Phys.}\ }\textbf {\bibinfo {volume} {7}},\ \bibinfo {pages} {52}
  (\bibinfo {year} {2010})}\BibitemShut {NoStop}%
\bibitem [{\citenamefont {Cummer}\ and\ \citenamefont
  {Schurig}(2007)}]{cummer2007one}%
  \BibitemOpen
  \bibfield  {author} {\bibinfo {author} {\bibfnamefont {S.~A.}\ \bibnamefont
  {Cummer}}\ and\ \bibinfo {author} {\bibfnamefont {D.}~\bibnamefont
  {Schurig}},\ }\href@noop {} {\bibfield  {journal} {\bibinfo  {journal} {New
  J. Phys.}\ }\textbf {\bibinfo {volume} {9}},\ \bibinfo {pages} {45} (\bibinfo
  {year} {2007})}\BibitemShut {NoStop}%
\bibitem [{\citenamefont {Torrent}\ and\ \citenamefont
  {S{\'a}nchez-Dehesa}(2007)}]{torrent2007acoustic}%
  \BibitemOpen
  \bibfield  {author} {\bibinfo {author} {\bibfnamefont {D.}~\bibnamefont
  {Torrent}}\ and\ \bibinfo {author} {\bibfnamefont {J.}~\bibnamefont
  {S{\'a}nchez-Dehesa}},\ }\href@noop {} {\bibfield  {journal} {\bibinfo
  {journal} {New J. Phys.}\ }\textbf {\bibinfo {volume} {9}},\ \bibinfo {pages}
  {323} (\bibinfo {year} {2007})}\BibitemShut {NoStop}%
\bibitem [{\citenamefont {McGaughey}\ \emph {et~al.}(2006)\citenamefont
  {McGaughey}, \citenamefont {Hussein}, \citenamefont {Landry}, \citenamefont
  {Kaviany},\ and\ \citenamefont {Hulbert}}]{McGaughey_2006}%
  \BibitemOpen
  \bibfield  {author} {\bibinfo {author} {\bibfnamefont {A.~J.~H.}\
  \bibnamefont {McGaughey}}, \bibinfo {author} {\bibfnamefont {M.~I.}\
  \bibnamefont {Hussein}}, \bibinfo {author} {\bibfnamefont {E.~S.}\
  \bibnamefont {Landry}}, \bibinfo {author} {\bibfnamefont {M.}~\bibnamefont
  {Kaviany}}, \ and\ \bibinfo {author} {\bibfnamefont {G.~M.}\ \bibnamefont
  {Hulbert}},\ }\href@noop {} {\bibfield  {journal} {\bibinfo  {journal} {Phys.
  Rev. B}\ }\textbf {\bibinfo {volume} {74}},\ \bibinfo {pages} {104304}
  (\bibinfo {year} {2006})}\BibitemShut {NoStop}%
\bibitem [{\citenamefont {Davis}\ and\ \citenamefont
  {Hussein}(2011)}]{davis2011thermal}%
  \BibitemOpen
  \bibfield  {author} {\bibinfo {author} {\bibfnamefont {B.~L.}\ \bibnamefont
  {Davis}}\ and\ \bibinfo {author} {\bibfnamefont {M.~I.}\ \bibnamefont
  {Hussein}},\ }\href@noop {} {\bibfield  {journal} {\bibinfo  {journal} {AIP
  Adv.}\ }\textbf {\bibinfo {volume} {1}},\ \bibinfo {pages} {041701} (\bibinfo
  {year} {2011})}\BibitemShut {NoStop}%
\bibitem [{\citenamefont {Davis}\ and\ \citenamefont
  {Hussein}(2014)}]{PhysRevLett.112.055505}%
  \BibitemOpen
  \bibfield  {author} {\bibinfo {author} {\bibfnamefont {B.~L.}\ \bibnamefont
  {Davis}}\ and\ \bibinfo {author} {\bibfnamefont {M.~I.}\ \bibnamefont
  {Hussein}},\ }\href {\doibase 10.1103/PhysRevLett.112.055505} {\bibfield
  {journal} {\bibinfo  {journal} {Phys. Rev. Lett.}\ }\textbf {\bibinfo
  {volume} {112}},\ \bibinfo {pages} {055505} (\bibinfo {year}
  {2014})}\BibitemShut {NoStop}%
\bibitem [{\citenamefont {Li}\ \emph {et~al.}(2012)\citenamefont {Li},
  \citenamefont {Ren}, \citenamefont {Wang}, \citenamefont {Zhang},
  \citenamefont {H{\"a}nggi},\ and\ \citenamefont {Li}}]{li2012colloquium}%
  \BibitemOpen
  \bibfield  {author} {\bibinfo {author} {\bibfnamefont {N.}~\bibnamefont
  {Li}}, \bibinfo {author} {\bibfnamefont {J.}~\bibnamefont {Ren}}, \bibinfo
  {author} {\bibfnamefont {L.}~\bibnamefont {Wang}}, \bibinfo {author}
  {\bibfnamefont {G.}~\bibnamefont {Zhang}}, \bibinfo {author} {\bibfnamefont
  {P.}~\bibnamefont {H{\"a}nggi}}, \ and\ \bibinfo {author} {\bibfnamefont
  {B.}~\bibnamefont {Li}},\ }\href@noop {} {\bibfield  {journal} {\bibinfo
  {journal} {Rev. Mod. Phys.}\ }\textbf {\bibinfo {volume} {84}},\ \bibinfo
  {pages} {1045} (\bibinfo {year} {2012})}\BibitemShut {NoStop}%
\bibitem [{\citenamefont {Deymier}(2013)}]{Deymier2013}%
  \BibitemOpen
  \bibfield  {author} {\bibinfo {author} {\bibfnamefont {P.~A.}\ \bibnamefont
  {Deymier}},\ }\href@noop {} {\emph {\bibinfo {title} {Acoustic Metamaterials
  and Phononic Crystals}}}\ (\bibinfo  {publisher} {Berlin Heidelberg:
  Springer-Verlag},\ \bibinfo {year} {2013})\BibitemShut {NoStop}%
\bibitem [{\citenamefont {Hussein}\ \emph {et~al.}(2014)\citenamefont
  {Hussein}, \citenamefont {Leamy},\ and\ \citenamefont
  {Ruzzene}}]{Hussein_AMR_2014}%
  \BibitemOpen
  \bibfield  {author} {\bibinfo {author} {\bibfnamefont {M.~I.}\ \bibnamefont
  {Hussein}}, \bibinfo {author} {\bibfnamefont {M.~J.}\ \bibnamefont {Leamy}},
  \ and\ \bibinfo {author} {\bibfnamefont {M.}~\bibnamefont {Ruzzene}},\
  }\href@noop {} {\bibfield  {journal} {\bibinfo  {journal} {Appl. Mech. Rev.}\
  }\textbf {\bibinfo {volume} {66}},\ \bibinfo {pages} {040802} (\bibinfo
  {year} {2014})}\BibitemShut {NoStop}%
\bibitem [{\citenamefont {Hussein}\ and\ \citenamefont
  {El-Kady}(2011)}]{hussein2011preface}%
  \BibitemOpen
  \bibfield  {author} {\bibinfo {author} {\bibfnamefont {M.~I.}\ \bibnamefont
  {Hussein}}\ and\ \bibinfo {author} {\bibfnamefont {I.}~\bibnamefont
  {El-Kady}},\ }\href@noop {} {\bibfield  {journal} {\bibinfo  {journal} {AIP
  Adv.}\ }\textbf {\bibinfo {volume} {1}},\ \bibinfo {pages} {041301} (\bibinfo
  {year} {2011})}\BibitemShut {NoStop}%
\bibitem [{\citenamefont {Hussein}\ \emph {et~al.}(2013)\citenamefont
  {Hussein}, \citenamefont {Leamy},\ and\ \citenamefont
  {Ruzzene}}]{hussein_JVA_2013}%
  \BibitemOpen
  \bibfield  {author} {\bibinfo {author} {\bibfnamefont {M.~I.}\ \bibnamefont
  {Hussein}}, \bibinfo {author} {\bibfnamefont {M.~J.}\ \bibnamefont {Leamy}},
  \ and\ \bibinfo {author} {\bibfnamefont {M.}~\bibnamefont {Ruzzene}},\
  }\href@noop {} {\bibfield  {journal} {\bibinfo  {journal} {J. Vib. Acoust.}\
  }\textbf {\bibinfo {volume} {135}},\ \bibinfo {pages} {040201} (\bibinfo
  {year} {2013})}\BibitemShut {NoStop}%
\bibitem [{\citenamefont {Graff}(1991)}]{graff1975wave}%
  \BibitemOpen
  \bibfield  {author} {\bibinfo {author} {\bibfnamefont {K.~F.}\ \bibnamefont
  {Graff}},\ }\href@noop {} {\emph {\bibinfo {title} {Wave Motion in Elastic
  Solids}}}\ (\bibinfo  {publisher} {Dover Publications},\ \bibinfo {year}
  {1991})\BibitemShut {NoStop}%
\bibitem [{\citenamefont {Achenbach}(1984)}]{achenbach1984wave}%
  \BibitemOpen
  \bibfield  {author} {\bibinfo {author} {\bibfnamefont {J.}~\bibnamefont
  {Achenbach}},\ }\href@noop {} {\emph {\bibinfo {title} {Wave Propagation in
  Elastic Solids}}}\ (\bibinfo  {publisher} {Elsevier},\ \bibinfo {year}
  {1984})\BibitemShut {NoStop}%
\bibitem [{\citenamefont {Bhatnagar}(1979)}]{bhatnagar1979nonlinear}%
  \BibitemOpen
  \bibfield  {author} {\bibinfo {author} {\bibfnamefont {P.~L.}\ \bibnamefont
  {Bhatnagar}},\ }\href@noop {} {\emph {\bibinfo {title} {Nonlinear Waves in
  One-dimensional Dispersive Systems}}},\ Vol.\ \bibinfo {volume} {142}\
  (\bibinfo  {publisher} {Clarendon Press Oxford},\ \bibinfo {year}
  {1979})\BibitemShut {NoStop}%
\bibitem [{\citenamefont {Ogden}(1997)}]{ogden1997non}%
  \BibitemOpen
  \bibfield  {author} {\bibinfo {author} {\bibfnamefont {R.~W.}\ \bibnamefont
  {Ogden}},\ }\href@noop {} {\emph {\bibinfo {title} {Non-linear Elastic
  Deformations}}}\ (\bibinfo  {publisher} {Courier Dover Publications},\
  \bibinfo {year} {1997})\BibitemShut {NoStop}%
\bibitem [{\citenamefont {Norris}(1998)}]{Norris_1998}%
  \BibitemOpen
  \bibfield  {author} {\bibinfo {author} {\bibfnamefont {A.~N.}\ \bibnamefont
  {Norris}},\ }\href@noop {} {\bibfield  {journal} {\bibinfo  {journal} {In: M.
  F. Hamilton and D. T. Blackstock (Eds.), Nonlinear Acoustics, 263--277,
  Academic Press, San Diego}\ } (\bibinfo {year} {1998})}\BibitemShut {NoStop}%
\bibitem [{\citenamefont {Porubov}(2003)}]{porubov2003amplification}%
  \BibitemOpen
  \bibfield  {author} {\bibinfo {author} {\bibfnamefont {A.}~\bibnamefont
  {Porubov}},\ }\href@noop {} {\emph {\bibinfo {title} {Amplification of
  Nonlinear Strain Waves in Solids}}}\ (\bibinfo  {publisher} {World
  Scientific},\ \bibinfo {year} {2003})\BibitemShut {NoStop}%
\bibitem [{\citenamefont {Erofeyev}(2003)}]{erofeyev2003wave}%
  \BibitemOpen
  \bibfield  {author} {\bibinfo {author} {\bibfnamefont {V.~I.}\ \bibnamefont
  {Erofeyev}},\ }\href@noop {} {\emph {\bibinfo {title} {Wave Processes in
  Solids with Microstructure}}},\ Vol.~\bibinfo {volume} {8}\ (\bibinfo
  {publisher} {World Scientific},\ \bibinfo {year} {2003})\BibitemShut
  {NoStop}%
\bibitem [{\citenamefont {Abedinnasab}\ and\ \citenamefont
  {Hussein}(2013)}]{abedinnasab2013wave}%
  \BibitemOpen
  \bibfield  {author} {\bibinfo {author} {\bibfnamefont {M.~H.}\ \bibnamefont
  {Abedinnasab}}\ and\ \bibinfo {author} {\bibfnamefont {M.~I.}\ \bibnamefont
  {Hussein}},\ }\href {\doibase
  http://dx.doi.org/10.1016/j.wavemoti.2012.10.008} {\bibfield  {journal}
  {\bibinfo  {journal} {Wave Motion}\ }\textbf {\bibinfo {volume} {50}},\
  \bibinfo {pages} {374 } (\bibinfo {year} {2013})}\BibitemShut {NoStop}%
\bibitem [{\citenamefont {Lee}\ and\ \citenamefont
  {Cai}(2013)}]{Lee_PNAS_2013}%
  \BibitemOpen
  \bibfield  {author} {\bibinfo {author} {\bibfnamefont {K.~G.}\ \bibnamefont
  {Lee}, \bibfnamefont {W.}}\ and\ \bibinfo {author} {\bibfnamefont
  {D.}~\bibnamefont {Cai}},\ }\href@noop {} {\bibfield  {journal} {\bibinfo
  {journal} {P. Natl. Acad. Sci. USA}\ }\textbf {\bibinfo {volume} {110}},\
  \bibinfo {pages} {3237–} (\bibinfo {year} {2013})}\BibitemShut {NoStop}%
\bibitem [{\citenamefont {Vakakis}\ and\ \citenamefont
  {King}(1995)}]{vakakis1995nonlinear}%
  \BibitemOpen
  \bibfield  {author} {\bibinfo {author} {\bibfnamefont {A.~F.}\ \bibnamefont
  {Vakakis}}\ and\ \bibinfo {author} {\bibfnamefont {M.~E.}\ \bibnamefont
  {King}},\ }\href@noop {} {\bibfield  {journal} {\bibinfo  {journal} {J.
  Acoust. Soc. Am.}\ }\textbf {\bibinfo {volume} {98}},\ \bibinfo {pages}
  {1534} (\bibinfo {year} {1995})}\BibitemShut {NoStop}%
\bibitem [{\citenamefont {Manktelow}\ \emph {et~al.}(2011)\citenamefont
  {Manktelow}, \citenamefont {Leamy},\ and\ \citenamefont
  {Ruzzene}}]{manktelow2011multiple}%
  \BibitemOpen
  \bibfield  {author} {\bibinfo {author} {\bibfnamefont {K.}~\bibnamefont
  {Manktelow}}, \bibinfo {author} {\bibfnamefont {M.~J.}\ \bibnamefont
  {Leamy}}, \ and\ \bibinfo {author} {\bibfnamefont {M.}~\bibnamefont
  {Ruzzene}},\ }\href@noop {} {\bibfield  {journal} {\bibinfo  {journal}
  {Nonlinear Dynam.}\ }\textbf {\bibinfo {volume} {63}},\ \bibinfo {pages}
  {193} (\bibinfo {year} {2011})}\BibitemShut {NoStop}%
\bibitem [{\citenamefont {Swinteck}\ \emph {et~al.}(2013)\citenamefont
  {Swinteck}, \citenamefont {Muralidharan},\ and\ \citenamefont
  {Deymier}}]{Swinteck_2013_JVA}%
  \BibitemOpen
  \bibfield  {author} {\bibinfo {author} {\bibfnamefont {N.~Z.}\ \bibnamefont
  {Swinteck}}, \bibinfo {author} {\bibfnamefont {K.}~\bibnamefont
  {Muralidharan}}, \ and\ \bibinfo {author} {\bibfnamefont {P.~A.}\
  \bibnamefont {Deymier}},\ }\href@noop {} {\bibfield  {journal} {\bibinfo
  {journal} {J. Vib. Acoust.}\ }\textbf {\bibinfo {volume} {135}},\ \bibinfo
  {pages} {041016} (\bibinfo {year} {2013})}\BibitemShut {NoStop}%
\bibitem [{\citenamefont {Chakraborty}\ and\ \citenamefont
  {Mallik}(2001)}]{chakraborty2001dynamics}%
  \BibitemOpen
  \bibfield  {author} {\bibinfo {author} {\bibfnamefont {G.}~\bibnamefont
  {Chakraborty}}\ and\ \bibinfo {author} {\bibfnamefont {A.}~\bibnamefont
  {Mallik}},\ }\href@noop {} {\bibfield  {journal} {\bibinfo  {journal} {Int.
  J. Nonlinear Mech.}\ }\textbf {\bibinfo {volume} {36}},\ \bibinfo {pages}
  {375} (\bibinfo {year} {2001})}\BibitemShut {NoStop}%
\bibitem [{\citenamefont {Narisetti}\ \emph {et~al.}(2010)\citenamefont
  {Narisetti}, \citenamefont {Leamy},\ and\ \citenamefont
  {Ruzzene}}]{narisetti2010perturbation}%
  \BibitemOpen
  \bibfield  {author} {\bibinfo {author} {\bibfnamefont {R.~K.}\ \bibnamefont
  {Narisetti}}, \bibinfo {author} {\bibfnamefont {M.~J.}\ \bibnamefont
  {Leamy}}, \ and\ \bibinfo {author} {\bibfnamefont {M.}~\bibnamefont
  {Ruzzene}},\ }\href@noop {} {\bibfield  {journal} {\bibinfo  {journal} {J.
  Vib. Acous.}\ }\textbf {\bibinfo {volume} {132}},\ \bibinfo {pages} {031001}
  (\bibinfo {year} {2010})}\BibitemShut {NoStop}%
\bibitem [{\citenamefont {Lazarov}\ and\ \citenamefont
  {Jensen}(2007)}]{lazarov2007low}%
  \BibitemOpen
  \bibfield  {author} {\bibinfo {author} {\bibfnamefont {B.~S.}\ \bibnamefont
  {Lazarov}}\ and\ \bibinfo {author} {\bibfnamefont {J.~S.}\ \bibnamefont
  {Jensen}},\ }\href@noop {} {\bibfield  {journal} {\bibinfo  {journal} {Int.
  J. Nonlinear Mech.}\ }\textbf {\bibinfo {volume} {42}},\ \bibinfo {pages}
  {1186} (\bibinfo {year} {2007})}\BibitemShut {NoStop}%
\bibitem [{\citenamefont {Narisetti}\ \emph {et~al.}(2012)\citenamefont
  {Narisetti}, \citenamefont {Ruzzene},\ and\ \citenamefont
  {Leamy}}]{narisetti2012study}%
  \BibitemOpen
  \bibfield  {author} {\bibinfo {author} {\bibfnamefont {R.~K.}\ \bibnamefont
  {Narisetti}}, \bibinfo {author} {\bibfnamefont {M.}~\bibnamefont {Ruzzene}},
  \ and\ \bibinfo {author} {\bibfnamefont {M.~J.}\ \bibnamefont {Leamy}},\
  }\href@noop {} {\bibfield  {journal} {\bibinfo  {journal} {Wave Motion}\
  }\textbf {\bibinfo {volume} {49}},\ \bibinfo {pages} {394} (\bibinfo {year}
  {2012})}\BibitemShut {NoStop}%
\bibitem [{\citenamefont {Manktelow}\ \emph {et~al.}(2013)\citenamefont
  {Manktelow}, \citenamefont {Leamy},\ and\ \citenamefont
  {Ruzzene}}]{manktelow2013comparison}%
  \BibitemOpen
  \bibfield  {author} {\bibinfo {author} {\bibfnamefont {K.}~\bibnamefont
  {Manktelow}}, \bibinfo {author} {\bibfnamefont {M.~J.}\ \bibnamefont
  {Leamy}}, \ and\ \bibinfo {author} {\bibfnamefont {M.}~\bibnamefont
  {Ruzzene}},\ }\href {\doibase
  http://dx.doi.org/10.1016/j.wavemoti.2012.12.009} {\bibfield  {journal}
  {\bibinfo  {journal} {Wave Motion}\ }\textbf {\bibinfo {volume} {50}},\
  \bibinfo {pages} {494 } (\bibinfo {year} {2013})}\BibitemShut {NoStop}%
\bibitem [{\citenamefont {Li}\ and\ \citenamefont {Li}(2012)}]{Li_2012}%
  \BibitemOpen
  \bibfield  {author} {\bibinfo {author} {\bibfnamefont {N.}~\bibnamefont
  {Li}}\ and\ \bibinfo {author} {\bibfnamefont {B.}~\bibnamefont {Li}},\
  }\href@noop {} {\bibfield  {journal} {\bibinfo  {journal} {AIP Adv.}\
  }\textbf {\bibinfo {volume} {2}},\ \bibinfo {pages} {041408} (\bibinfo {year}
  {2012})}\BibitemShut {NoStop}%
\bibitem [{\citenamefont {Andrianov}\ \emph {et~al.}(2013)\citenamefont
  {Andrianov}, \citenamefont {Danishevs'kyy}, \citenamefont {Ryzhkov},\ and\
  \citenamefont {Weichert}}]{Andrianov_WM_2013}%
  \BibitemOpen
  \bibfield  {author} {\bibinfo {author} {\bibfnamefont {I.~V.}\ \bibnamefont
  {Andrianov}}, \bibinfo {author} {\bibfnamefont {V.~V.}\ \bibnamefont
  {Danishevs'kyy}}, \bibinfo {author} {\bibfnamefont {O.~I.}\ \bibnamefont
  {Ryzhkov}}, \ and\ \bibinfo {author} {\bibfnamefont {D.}~\bibnamefont
  {Weichert}},\ }\href@noop {} {\bibfield  {journal} {\bibinfo  {journal} {Wave
  Motion}\ }\textbf {\bibinfo {volume} {50}},\ \bibinfo {pages} {271–281}
  (\bibinfo {year} {2013})}\BibitemShut {NoStop}%
\bibitem [{\citenamefont {Daraio}\ \emph
  {et~al.}(2006{\natexlab{a}})\citenamefont {Daraio}, \citenamefont
  {Nesterenko}, \citenamefont {Herbold},\ and\ \citenamefont
  {Jin}}]{daraio2006tunability}%
  \BibitemOpen
  \bibfield  {author} {\bibinfo {author} {\bibfnamefont {C.}~\bibnamefont
  {Daraio}}, \bibinfo {author} {\bibfnamefont {V.~F.}\ \bibnamefont
  {Nesterenko}}, \bibinfo {author} {\bibfnamefont {E.~B.}\ \bibnamefont
  {Herbold}}, \ and\ \bibinfo {author} {\bibfnamefont {S.}~\bibnamefont
  {Jin}},\ }\href@noop {} {\bibfield  {journal} {\bibinfo  {journal} {Phys.
  Rev. E}\ }\textbf {\bibinfo {volume} {73}},\ \bibinfo {pages} {026610}
  (\bibinfo {year} {2006}{\natexlab{a}})}\BibitemShut {NoStop}%
\bibitem [{\citenamefont {Herbold}\ \emph {et~al.}(2009)\citenamefont
  {Herbold}, \citenamefont {Kim}, \citenamefont {Nesterenko}, \citenamefont
  {Wang},\ and\ \citenamefont {Daraio}}]{herbold2009pulse}%
  \BibitemOpen
  \bibfield  {author} {\bibinfo {author} {\bibfnamefont {E.~B.}\ \bibnamefont
  {Herbold}}, \bibinfo {author} {\bibfnamefont {J.}~\bibnamefont {Kim}},
  \bibinfo {author} {\bibfnamefont {V.~F.}\ \bibnamefont {Nesterenko}},
  \bibinfo {author} {\bibfnamefont {S.~Y.}\ \bibnamefont {Wang}}, \ and\
  \bibinfo {author} {\bibfnamefont {C.}~\bibnamefont {Daraio}},\ }\href@noop {}
  {\bibfield  {journal} {\bibinfo  {journal} {Acta Mech.}\ }\textbf {\bibinfo
  {volume} {205}},\ \bibinfo {pages} {85} (\bibinfo {year} {2009})}\BibitemShut
  {NoStop}%
\bibitem [{\citenamefont {Daraio}\ \emph
  {et~al.}(2006{\natexlab{b}})\citenamefont {Daraio}, \citenamefont
  {V.~F.~Nesterenko}, \citenamefont {Herbold},\ and\ \citenamefont
  {Jin}}]{Daraio_2006_PRL}%
  \BibitemOpen
  \bibfield  {author} {\bibinfo {author} {\bibfnamefont {C.}~\bibnamefont
  {Daraio}}, \bibinfo {author} {\bibfnamefont {V.~F.}\ \bibnamefont
  {V.~F.~Nesterenko}}, \bibinfo {author} {\bibfnamefont {E.~B.}\ \bibnamefont
  {Herbold}}, \ and\ \bibinfo {author} {\bibfnamefont {S.}~\bibnamefont
  {Jin}},\ }\href@noop {} {\bibfield  {journal} {\bibinfo  {journal} {Phys.
  Rev. Lett.}\ }\textbf {\bibinfo {volume} {96}},\ \bibinfo {pages} {058002}
  (\bibinfo {year} {2006}{\natexlab{b}})}\BibitemShut {NoStop}%
\bibitem [{\citenamefont {Spadoni}\ and\ \citenamefont
  {Daraio}(2010)}]{Spadoni_2010_PNAS}%
  \BibitemOpen
  \bibfield  {author} {\bibinfo {author} {\bibfnamefont {A.}~\bibnamefont
  {Spadoni}}\ and\ \bibinfo {author} {\bibfnamefont {C.}~\bibnamefont
  {Daraio}},\ }\href@noop {} {\bibfield  {journal} {\bibinfo  {journal} {P.
  Natl. Acad. Sci. USA}\ }\textbf {\bibinfo {volume} {107}},\ \bibinfo {pages}
  {7230} (\bibinfo {year} {2010})}\BibitemShut {NoStop}%
\bibitem [{\citenamefont {Boechler}\ \emph {et~al.}(2011)\citenamefont
  {Boechler}, \citenamefont {Theocharis},\ and\ \citenamefont
  {Daraio}}]{Boechler_2011_NM}%
  \BibitemOpen
  \bibfield  {author} {\bibinfo {author} {\bibfnamefont {N.}~\bibnamefont
  {Boechler}}, \bibinfo {author} {\bibfnamefont {G.}~\bibnamefont
  {Theocharis}}, \ and\ \bibinfo {author} {\bibfnamefont {C.}~\bibnamefont
  {Daraio}},\ }\href@noop {} {\bibfield  {journal} {\bibinfo  {journal} {Nat.
  Mater.}\ }\textbf {\bibinfo {volume} {10}},\ \bibinfo {pages} {665} (\bibinfo
  {year} {2011})}\BibitemShut {NoStop}%
\bibitem [{Note1()}]{Note1}%
  \BibitemOpen
  \bibinfo {note} {A phononic material in general may be classified into two
  types, a phononic crystal and a locally resonant elastic metamaterial~\cite
  {Deymier2013,Hussein_AMR_2014}. In this work we focus on the former, but the
  mathematical treatment is also applicable to the latter~\cite
  {Khajehtourian_Hussein_2013}.}\BibitemShut {Stop}%
\bibitem [{\citenamefont {Bloch}(1929)}]{bloch1929quantenmechanik}%
  \BibitemOpen
  \bibfield  {author} {\bibinfo {author} {\bibfnamefont {F.}~\bibnamefont
  {Bloch}},\ }\href@noop {} {\bibfield  {journal} {\bibinfo  {journal} {Z.
  Phys.}\ }\textbf {\bibinfo {volume} {52}},\ \bibinfo {pages} {555} (\bibinfo
  {year} {1929})}\BibitemShut {NoStop}%
\bibitem [{\citenamefont {Hussein}\ \emph {et~al.}(2006)\citenamefont
  {Hussein}, \citenamefont {Hulbert},\ and\ \citenamefont
  {Scott}}]{hussein2006dispersive}%
  \BibitemOpen
  \bibfield  {author} {\bibinfo {author} {\bibfnamefont {M.~I.}\ \bibnamefont
  {Hussein}}, \bibinfo {author} {\bibfnamefont {G.~M.}\ \bibnamefont
  {Hulbert}}, \ and\ \bibinfo {author} {\bibfnamefont {R.~A.}\ \bibnamefont
  {Scott}},\ }\href@noop {} {\bibfield  {journal} {\bibinfo  {journal} {J.
  Sound Vib.}\ }\textbf {\bibinfo {volume} {289}},\ \bibinfo {pages} {779}
  (\bibinfo {year} {2006})}\BibitemShut {NoStop}%
\bibitem [{\citenamefont {Billingham}\ and\ \citenamefont
  {King}(2000)}]{billingham2000wave}%
  \BibitemOpen
  \bibfield  {author} {\bibinfo {author} {\bibfnamefont {J.}~\bibnamefont
  {Billingham}}\ and\ \bibinfo {author} {\bibfnamefont {A.~C.}\ \bibnamefont
  {King}},\ }\href@noop {} {\emph {\bibinfo {title} {Wave motion}}},\ \bibinfo
  {number} {24}\ (\bibinfo  {publisher} {Cambridge University Press},\ \bibinfo
  {year} {2000})\BibitemShut {NoStop}%
\bibitem [{\citenamefont {Nayfeh}\ and\ \citenamefont
  {Mook}(2008)}]{nayfeh2008nonlinear}%
  \BibitemOpen
  \bibfield  {author} {\bibinfo {author} {\bibfnamefont {A.~H.}\ \bibnamefont
  {Nayfeh}}\ and\ \bibinfo {author} {\bibfnamefont {D.~T.}\ \bibnamefont
  {Mook}},\ }\href@noop {} {\emph {\bibinfo {title} {Nonlinear Oscillations}}}\
  (\bibinfo  {publisher} {Wiley},\ \bibinfo {year} {2008})\BibitemShut
  {NoStop}%
\bibitem [{\citenamefont {Khajehtourian}\ and\ \citenamefont
  {Hussein}(2013)}]{Khajehtourian_Hussein_2013}%
  \BibitemOpen
  \bibfield  {author} {\bibinfo {author} {\bibfnamefont {R.}~\bibnamefont
  {Khajehtourian}}\ and\ \bibinfo {author} {\bibfnamefont {M.~I.}\ \bibnamefont
  {Hussein}},\ }\href@noop {} {\bibfield  {journal} {\bibinfo  {journal}
  {Proceedings of Phononics 2013, Paper PHONONICS-2013-0175, pp. 180-181}\ }
  (\bibinfo {year} {2013})}\BibitemShut {NoStop}%
\end{thebibliography}%
\end{document}